\input{aipcheck}

\documentclass[
    ,final            % use final for the camera ready runs
  ]
  {aipproc}

\layoutstyle{6x9}

\begin{document}

\title{Understanding the baryon and meson spectra}


\author{M.R. Pennington}{
  address={Theory Center, Jefferson Lab, 12000 Jefferson Avenue, Newport News, VA 23606, U.S.A.}
}

\begin{abstract}
A brief overview is given of what we know of the baryon and meson spectra, with a focus on what are the key internal degrees of freedom and how these relate to strong coupling QCD. The challenges, experimental, theoretical and phenomenological, for the future are outlined, with particular reference to a program at Jefferson Lab to extract hadronic states in which glue unambiguously contributes to their quantum numbers.

\end{abstract}

\maketitle

%%%%%%%%%%%%%%%%%%%%%%%%%%%%%%%%%%%%%%%%%%%%
%% MAINMATTER
%%%%%%%%%%%%%%%%%%%%%%%%%%%%%%%%%%%%%%%%%%%%

\section{Revealing the workings of strong QCD}

With eyes fixed on the wonders of the LHC at the TeV scale, one may question why is the physics of the strong interaction at  1 GeV of  interest any longer. Is this not all ancient history? However, it is at the GeV scale that we already know the scalar sector that gives mass to most of  the visible universe. A GeV  is the energy scale at which we have discovered half the particles of a possible supersymmetric world. New strong interactions may await discovery, but QCD is the only strong interaction we already know. We should study it in as much detail as we can. After all it determines the properties of the nuclear matter of which we are made.  It is the strength of this interaction that brings a complexity of phenomena that outshines those of  perturbative electroweak physics. The richness of the tapestry of strong QCD is to be seen in the hadrons, their properties and structure, that it creates.

The paradigm for what can be learnt from spectroscopy is provided by atoms. Even if we did not have enough energy to separate electrons from the nucleus, we would know by studying the spectrum that though atoms are electrically neutral, they behave as though they are made  of electrically charged objects held together by an electromagnetic force governed by the rules of Quantum Electrodynamics. In a similar way color neutral hadrons are built of constituents carrying color charge, bound by the rules of QCD. But what are these rules? While asymptotic freedom provides a well exploited simplification for hard scattering processes, it is the fact that  over a distance of a fermi the interaction is strong that makes QCD so challenging  and why we look to experiment for guidance on how it really works. Strong coupling confines quarks and breaks chiral symmetry, and so defines the world of light hadrons. Quark confinement
is reflected in the spectrum and properties of hadrons, and we can learn from what experiment teaches about these.
% While for heavy hadrons of  bottom flavor (and even charmed quarks) ideas of potential theory are borne out by deeply bound state below the ${\overline B}B$ (and ${\overline D}D$) threshold, such non-relativistic notions are irrelevant to light flavored hadrons. 
We ask: what are the internal degrees of freedom of hadron states? The quark model, that was of course the seed from which the idea of QCD first germinated, suggests these are  constituent quarks (and anti-quarks). But is that all there are? What is the role of glue? Do gluons just stick the quarks together, and nothing more?

It is in the spectrum of charmonium that we have a working template from which to judge complexity most readily. Below ${\overline D}D$ threshold it all appears simple. We have the tightly bound systems of $J/\psi$, $\psi'$, $\eta_c \, \cdots$, as given by  non-relativistic potential models. Above the open charm threshold, we once thought the (almost) stable charmonia are replaced by states with 1-50 MeV widths decaying to ${\overline D}D$, ${\overline D}D^*$, ${\overline D^*}D^*$, ${\overline D_s}D_s^*, \cdots$, as their mass increases. What we find is that the states predicted by potential models are shifted by tens of MeV themselves: the decays affect their dynamics~\cite{barnes-swanson,wilson}. Hadronic decay channels are an essential degree of freedom. These not only shift predominantly ${\overline c}c$ states, but generate states that would not have existed without these hadron channels. The first discovery of a state of this type is the $X(3872)$, whose very existence is tied to the dynamics of the ${\overline D}^0 D^{*0}$ channel~\cite{tornqvist}.  More new states, a string of $X,\,Y$ and $Z$ states perhaps only exist because of their hadronic decays, sometimes these channels  binding in molecular (or multiquark) configurations. As dynamically coupled channel models have long suggested~\cite{vanbev-lutz}, hadrons and their decays are intimately related. Only for ground states may one think of them as having minimal quark configurations.

\section{What are the degrees of freedom in each hadron?}

Baryons have a special place
in the study of hadrons, as their structure is most obviously related to the color degree of freedom.
While a color singlet quark-antiquark system is basically the same however many colors there are, the
minimum number of quarks in a baryon is intimately tied to the number of colors. If $N_c$ were some other number than 3, the world would 
be quite different. Recognizing the flavor pattern of the ground state baryons was the key step  in the development of the  quark model. Consequently, this  model
with three independent quark degrees of freedom~\cite{isgur,capstick} has naturally served as the paradigm for what we expect the spectrum of excited baryons, both nucleons and $\Delta$'s, to look like too. While  experiment has long confirmed the lower lying states, many of the heavier ones  seemed to be {\it missing} above 1.6 GeV. 
 
If baryons were diquark--quark systems, as noted more than 40 years ago~\cite{lichtenberg}, the number of states  would be restricted and in fact be very like that observed uptil a year or so ago. 
However most of the early evidence on the baryon spectrum was accumulated from $\pi N$ scattering, and decays into the same channel.
Perhaps the {\it missing} states are just {\it dark} in these channels, and \lq\lq shine'' most in $\pi\pi N$ and $KY$. Consequently, the experimental program has concentrated more recently on these channels,  which are an increasing part of the $\pi N$ total cross-section as the energy goes up. 

\begin{figure}[t]
  \includegraphics[height=.24\textheight]{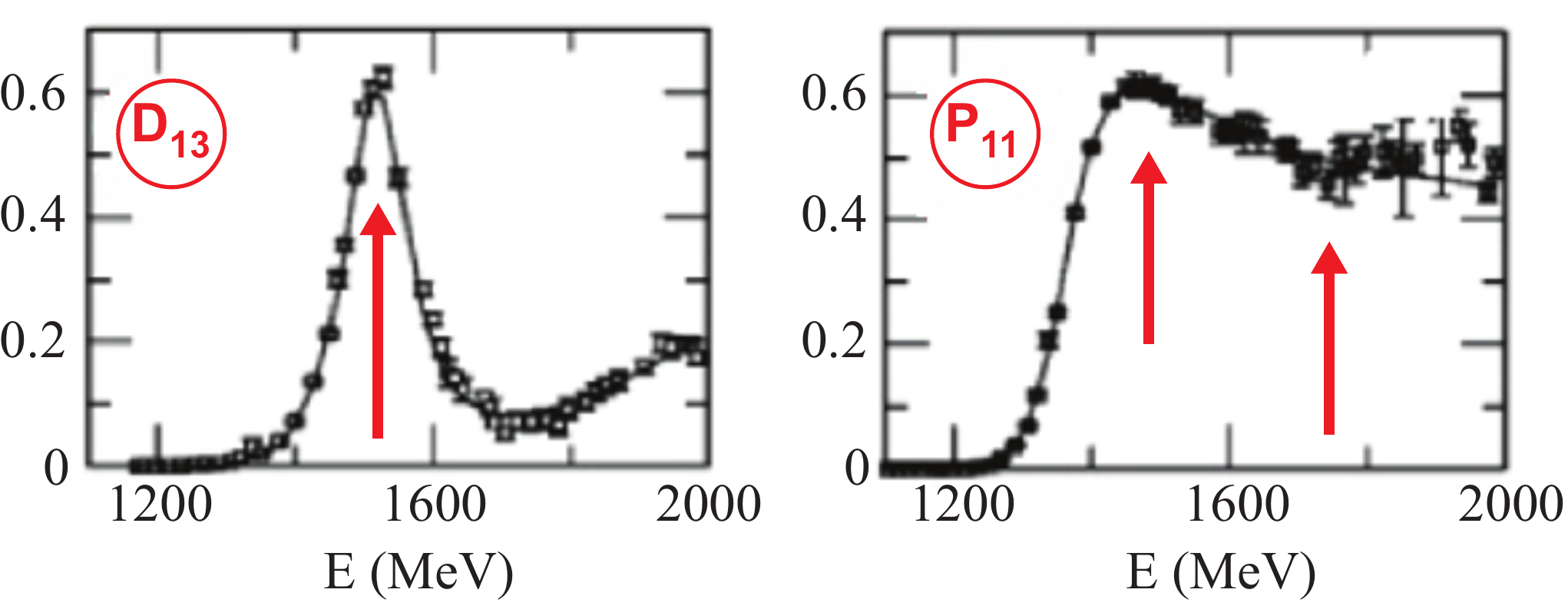}
\vspace{3mm}
  \caption{The imaginary parts of the $I=1/2$ $\pi N\to\pi N$ partial wave amplitudes, labeled by the quantum numbers $L_{2I\,2J}\,=\, D_{13}$ and $P_{11}$ from the SAID analysis~\cite{said} as functions of the $\pi N$ c.m. energy, $E$. The arrows mark the real part of the resonance pole positions.}
\vspace{-5mm}
\end{figure} 

But first how do we identify states in the spectrum of hadrons? Since states have definite quantum numbers, spin, parity, isospin etc, we have to decompose the observed data, integrated and differential cross-sections, into partial waves that specify these quantum numbers. To do this completely for processes with spin requires measurements with polarized beams and polarized targets. Having separated the partial waves, one finds it is only for the lowest mass state with a given quantum numbers that the partial wave looks anything like a simple Breit-Wigner resonance, see, as an example, the $D_{13}$ wave in Fig.~1~\cite{said}. Higher mass states are much less obvious. For instance in the $P_{11}$ wave of Fig.~1, while the $N^*(1440)$ (the Roper) appears as a bump in the imaginary part (and modulus), the higher mass $N^*(1710)$ can barely be seen in the same $\pi N\to\pi N$ amplitude. It is highly inelastic. A~state in the spectrum is then only identifiable by its pole in the complex energy plane on some  nearby unphysical sheet.
It is the poles that are the universal outcome of any modern amplitude analysis, as recognized by the PDG~\cite{pdg2012}.
 
By now a vast amount of data has been accumulated, and is being accumulated, on a wide range of baryonic processes, most recently initiated by real and virtual photon beams. The presence of many decay channels and the large widths to each of these demands coupled-channel amplitude analyses be performed. This requires a rich supply of input data if the richness of the spectrum is to be exposed.
 Thus from JLab~\cite{jlab-photo,jlab-data} and from ELSA~\cite{elsa-photo}, we have thousands of data on $\gamma p\to \pi^0 p$ and $KY$,  differential cross-sections and polarizations. These feature prominently in the latest analyses. 

The most ambitious analysis is that by the Excited Baryon Analysis Center (EBAC) team led by Harry Lee~\cite{ebac}. Not only does this fit a very wide range of data on baryonic channels, but it does this in terms of an effective field theory of hadronic interactions developed by Sato and Lee~\cite{ebac}. Their calculational procedure ensures unitarity is fulfilled, and  their Lagrangian provides a framework in which to consider the  nature and structure of each resonant state, and its \lq\lq core'' revealed. \lq\lq Bare'' or \lq\lq core'' states are those with no decays~\cite{ebacpoles}. While  for heavy quark systems one might reasonably define such bare states as those that arise in a potential model for charmonium or bottomonium, for light quark systems the model template is not so obvious. Here it is the Sato-Lee Lagrangian. How are such \lq\lq bare'' states connected to QCD? In fact are these connected to QCD at all? Perhaps there is no limit of QCD in which the hadronic decays of bound states can be turned off. Notwithstanding such interpretations, the results for the $N^*$ and $\Delta^*$ spectra of EBAC up to 1800 MeV have now been finalized~\cite{ebac2}, and are shown in Fig.~2. Their analysis of the detailed nature of these states is to come.

\begin{figure}[h]
  \includegraphics[height=.47\textheight]{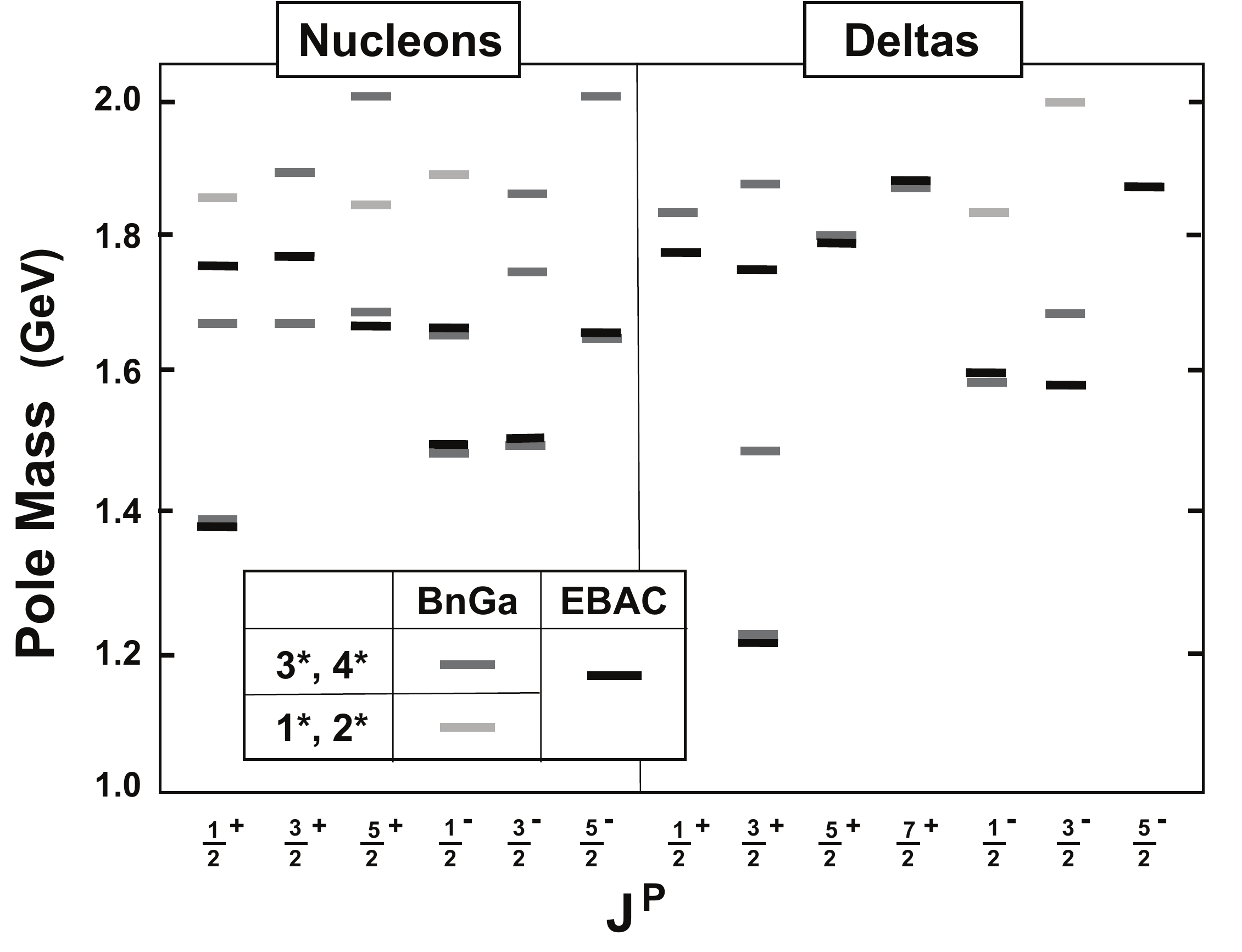}
\vspace{4mm}
  \caption{$N^*$ and $\Delta^*$ spectra, labeled by their spin and parity as $J^P$ along the abscissa, and the real part of the resonance pole positions along the ordinate, from the EBAC~\cite{ebac2} and Bonn-Gatchina~\cite{bn-ga} analyses. For the EBAC analysis all the states have $3^*-4^*$ provenance, while Bonn-Gatchina also include those with $1^*-2^*$ ratings, according to the legend shown. }
\end{figure}
A more  computationally flexible amplitude analysis program has been carried out by the Bonn-Gatchina team~\cite{sarantsev}. They fit an even more extensive range of multi-hadron final states and so are able to present results up to a higher energy~\cite{bn-ga}. Their states up to 2.1 GeV are shown in Fig.~2 too, with their assignment of their 1-4 star confidence~\cite{pdg2012}. The EBAC and Bonn-Gatchina spectra and couplings are very similar, but not identical. The larger mass range fitted includes the JLab data on channels such as $\gamma p\to K Y$~\cite{jlab-data} and this has enabled a number of the \lq\lq dark'' or \lq\lq missing'' baryons at last  to be revealed, like the $1/2^+$ $N^*(1880)$ and the $3/2^+$ $N^*(1900)$~\cite{bn-ga}.

Experiments on $\gamma p \to K^+ \Lambda$ with polarized beam and polarized target, together with the spin information from the weak decay $\Lambda \to \pi N$, allow more observables to be measured than the minimum needed to determine all the independent amplitudes (up to an overall phase)~\cite{tiator,sandorfi}. These {\it  over-complete} experiments hold out the prospect of checking that the partial wave solution that results in the spectrum shown in Fig.~2 is indeed the correct one. The development of polarized targets, such as FROST and HDice at JLab~\cite{jlab-targets}, have allowed neutron scattering data to be determined too. These results are eagerly awaited as they are an essential component in securing the partial wave solution and its isospin decomposition. 

Fig.~2 only shows the spectrum with zero strangeness. Within a simple quark model picture (which we have stressed may not be a realistic guide for highly excited states with their complex multi-hadron decays), baryons form flavor multiplets. Consequently, searching for baryons in the $\Sigma^*,\, \Lambda^*,\,\Xi^*$ families is a key part of the future experimental program. Such states have fewer (or better separated) hadronic decay channels and so may be narrower and more easily identifiable. 

\begin{figure}[th]
  \includegraphics[height=.28\textheight]{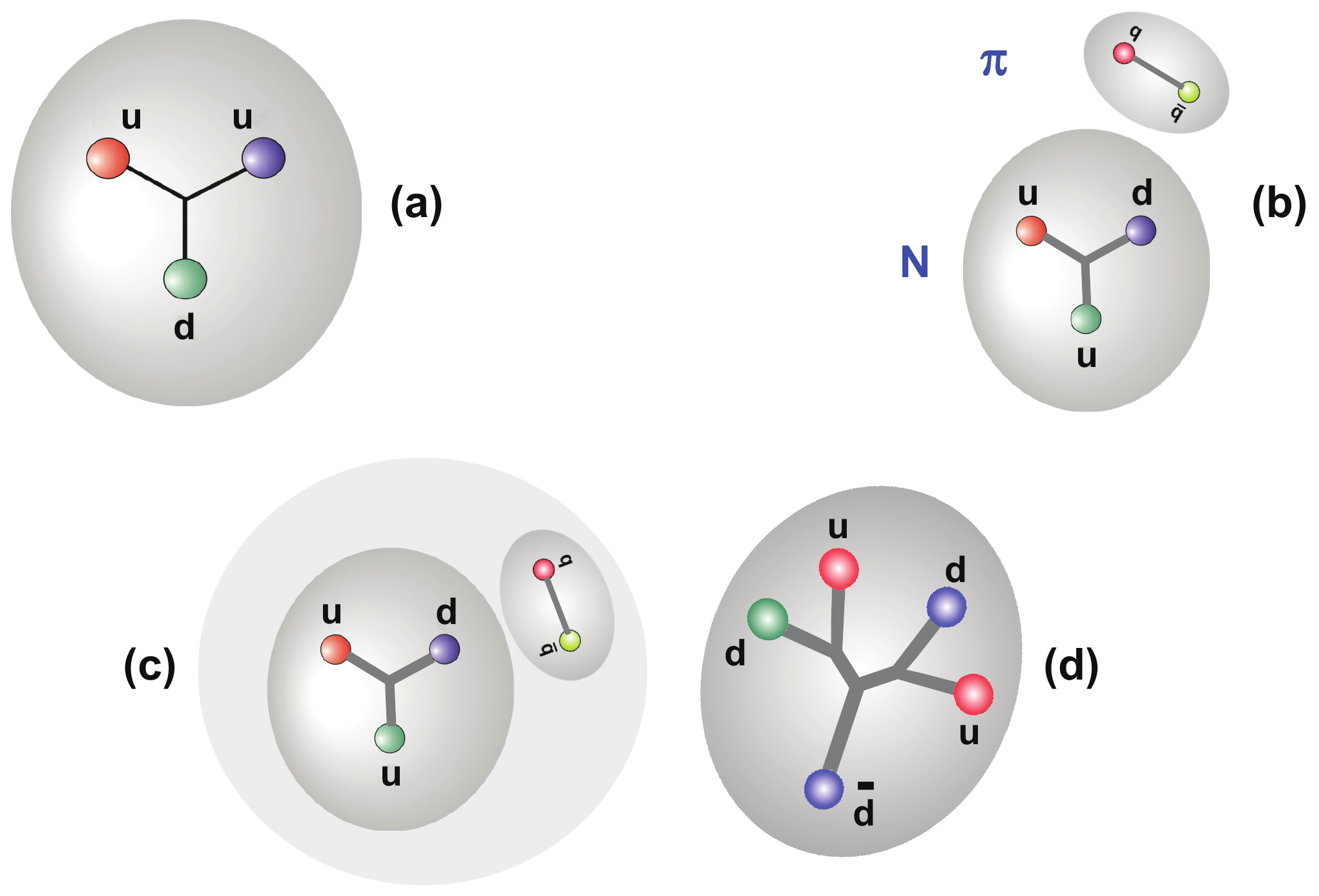}
  \caption{Cartoon of the possible Fock components (a-d) of some excited baryon, for instance the $N^*(1440)$. It almost certainly has components (a) and (b), but the relative amounts of (a-d) awaits to be determined for the Roper, or any other excited, baryon.}
\vspace{-5mm}
\end{figure}
 
Such results will teach us the Fock space decomposition of each resonant state. All but the ground states are inevitably complicated. As an example, the Roper, the $N^*(1440)$, cannot just be a three quark state, as depicted in Fig.~3a. It must have an explicit $\pi N$ component in its Fock space, Fig.~3b, since it is through this component (amongst others) that it decays. Its Fock space might then  be thought to include a nucleon and a pion (or even a multi-pion) cloud (Fig.~3c), but  might also contain a pentaquark configuration, like that in Fig.~3d.  Dynamical models, and eventually QCD, will tell us what are the proportion of these components for each physical state. Such compositions are also probed experimentally in  photo-transition processes. Once the data on these from the final running of CEBAF at 6 GeV are analyzed appropriately in terms of pole quantities~\cite{mokeev} we may have a better idea.

How is the spectrum of Fig.~2 related to QCD? The lattice provides the most consistent theoretical connection. The four-dimensional world is modeled as a discrete space-time to make the problem computationally feasible. The baryon spectrum computed most recently~\cite{edwards} reveals a pattern very like that of the quark model: certainly not that of a pointlike diquark--quark system. The \lq\lq missing'' states are there. However, one essential ingredient is clearly  {\it missing} in such
calculations. While continuum hadronic effects are included, they are not yet those of the physical world.  Though great computational strides have been made, the {\it up} and {\it down} quark masses are 8-15 times their physical value and so the pion mass is still 3 or 4 times too heavy. Consequently, the Fock space decomposition of the excited baryons is not physical. In terms of the pictures in Fig.~3,  components (b) and (c) are much much smaller than those of the real world, and so it's perhaps not surprising that the quark model-like Fig.~3a dominates. However, calculational progress towards a 140 MeV pion mass continues.

A continuum approach to QCD with  physical mass quarks  is  provided by the solution of the  Schwinger-Dyson/Bethe-Salpeter (SD/BS) system of equations~\cite{sdbsreviews}. There has been steady progress over decades in solving this complex system self-consistently. However, speedier computations are made possible by modeling the gluon by a simple contact interaction and presuming that baryons are bound states of a quark with an extended (not pointlike) diquark. Detailed calculations of the $N^*$ spectrum have then been made~\cite{cdr}. These include no decays and so no hadronic components. Amusingly there is a \lq\lq bare'' $P_{11}$ state that can be identified with the EBAC \lq\lq core'' state~\cite{ebacpoles}. The physical Roper is $\sim 500$ MeV lighter. As with the more ambitious SD/BS approach treating baryons as full three quark systems~\cite{eichmann}, these calculations must include decays if a meaningful comparison of  excited states with experiment is to be achieved. 

\section{Mesons: is this where glue is to be found?}
We now turn to mesons, first in the quark model.  The ${\overline q}q$ pair can have spin, $S_{qq}$, equal to 0 or 1. When combined with units of orbital angular momentum $L_{qq}$, they make a series of flavor multiplets, with each unit of $L_{qq}$ adding $\sim 700$ MeV of mass. The ground states with $L_{qq}=0$ have $J^{PC}\,=\,0^{-+}$ and $1^{--}$ quantum numbers. While the light pseudoscalars,  being the Goldstone bosons of chiral symmetry breaking, have atypical dynamics, the vector multiplet gives the ideally mixed paradigm, replicated by the mesons with higher $J=L_{qq}+1$.

The scalar ${\overline q}q$ multiplet is part of the $L_{qq}=S_{qq}=1$ family. There are at least 19 scalars below 2 GeV~\cite{pdg2012}, far more than can fit into one nonet~\cite{mrp-scalars}. It was Jaffe in his seminal work on multiquark states~\cite{jaffe} that recognized that the scalars below 1~GeV might be tetraquark states, while the more conventional ${\overline q}q$ $\,0^{++}$ mesons would be up close to their $2^{++}$ companions around 1.3 GeV. 
Such an interpretation naturally explains how the isosinglet $f_0(980)$ and isotriplet $a_0(980)$ can be degenerate in mass and both couple strongly to ${\overline K}K$: each is a $\overline{sn}sn$ state, with $n = u, d$. 
However, recent studies~\cite{menu2009,wilson2}, making use of the fine energy binning possible with BaBar data~\cite{marco}, have shown that the $f_0(980)$ is dominated by long range ${\overline K}K$ components, rather than a tighter bound tetraquark configuration. Similarly, the $\sigma$ and the $\kappa$ seem to be dominated by $\pi\pi$ and $\pi K$ components: their masses depending far more on their couplings to these channels than related to any simple quark mixing scheme. Indeed long ago, the dynamical calculation by van Beveren and Rupp~\cite{vanbev} highlighted how scalar ${\overline q}q$ seeds up at 1.3 GeV can give rise to two multiplets of hadrons, when their strong couplings to di-meson channels are included: an explicit example of dynamical coupled channel effects.   
 
\begin{figure}[h]
  \includegraphics[height=.45\textheight]{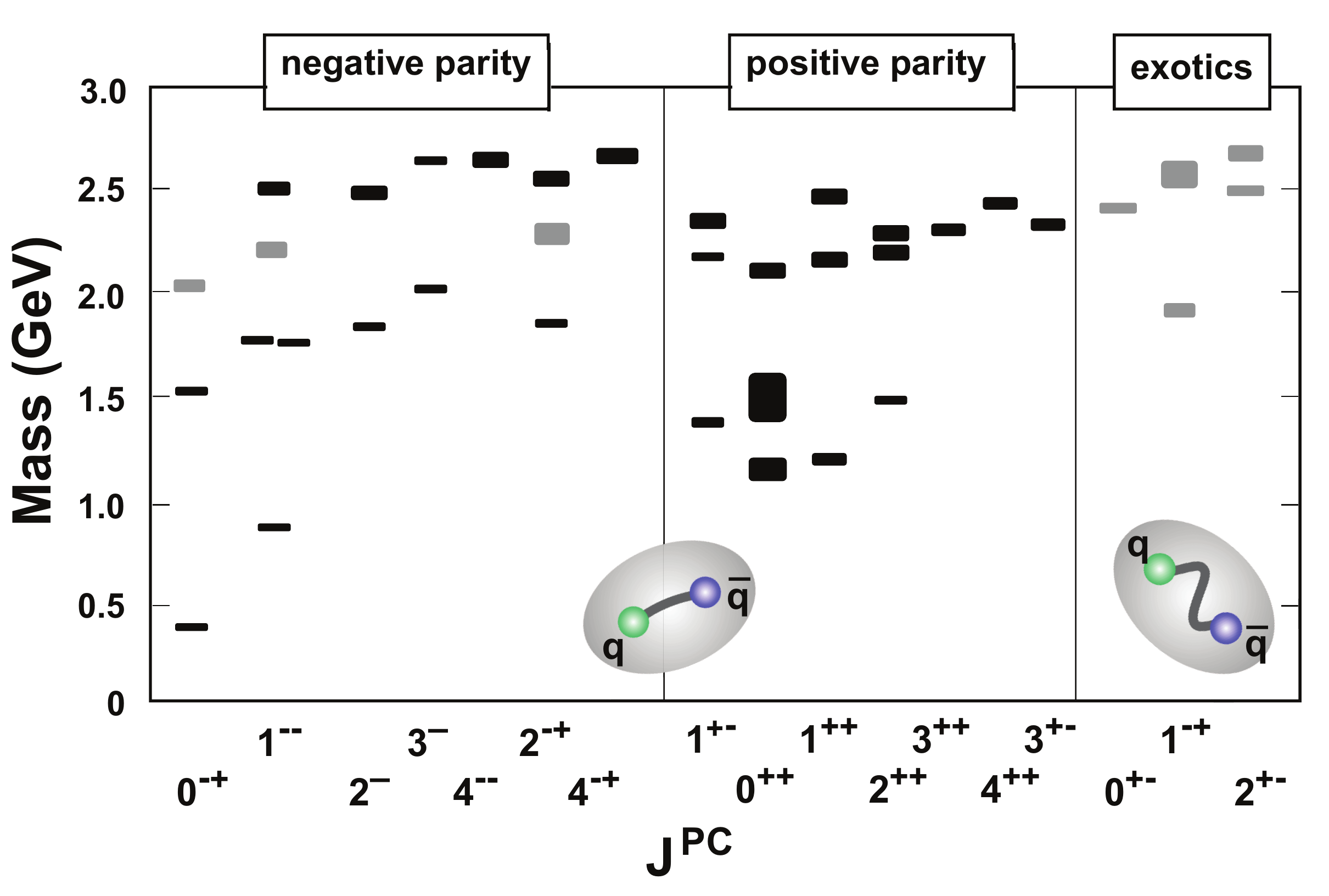}
  \caption{The isovector meson spectrum from the lattice calculations of Dudek {\it et al.}~\cite{dudek} with $m_\pi\,=\,396$ MeV, arranged according to their $J^{PC}$ quantum numbers.  Those found with ${\overline q}q$ operators are shown as black blocks, the size of which denote the statistical uncertainties. States from ${\overline q}qg$ operators are shown as grey blocks. Some of these have spin-exotic quantum numbers. These are shown on the right.}
\vspace{-5mm}
\end{figure}

Ever since the QCD Lagrangian was written down, it was recognized that there may exist hadrons with more complicated configurations than those of the simple quark model: states in which gluons pay a role in determining their quantum numbers. At first, calculations and experimental searches were for states made purely of glue. While many sightings were claimed, they never stood up to challenge~\cite{mrp-lund}. Indeed, it was quickly realized that any meson made of glue ({\it viz. glueballs}) must couple to quarks in order to decay into pions and kaons, and so mixing with these quark configurations is inevitable and could easily be large. Thus in the scalar sector discussed above,  several states between 1.3 and 1.8 GeV might have sizeable admixtures of glue, {\it viz.} $gg$,  without any being predominantly a glueball. That is a detail of dynamics that we do not yet understand, except in unrealistically simple mixing schemes. Consequently, attention has turned to other meson quantum numbers than those of the vacuum.

Lattice computations of the ${\overline q}q$ spectrum are approaching a maturity that includes all the states we know of from experiment, as shown in  first two columns of Fig.~4. There is displayed the results of the present state-of-the art computations for isovector mesons from Dudek {\it et al.}~\cite{dudek}. By using an inventive and ingenious set of operators, they have also been able to compute the spectrum of states that are ${\overline q}qg$. The grey blocks in Fig.~4 denote these hybrid states. On the left are seen hybrids with conventional quantum numbers, where exciting {\it glue} is found to require an extra $\sim 800$ MeV of mass. In addition, states with spin-exotic quantum numbers appear on the right of Fig.~4. The lightest is that with $J^{PC}=1^{-+}$, as long had been expected. At a pion mass of 400 MeV, this hybrid is found to be up around 2~GeV. Of course, a real mass pion is expected to affect this: in general making it lighter and broader. 

Possible states with  $1^{-+}$ quantum numbers were claimed in  a series of searches starting more than 35 years ago with GAMS~\cite{gams}, then (as shown in Fig.~5)  BNL-E852~\cite{chung} and VES~\cite{ves} ten years later. All find enhancements in the relevant partial wave. However, these signals only constitute a few percent of the integrated cross-section, and inevitably have $1^{-+}$ waves with sizeable uncertainties~\cite{dzierba}. Consequently, these experiments were never able to show that the underlying partial waves were resonant with a pole in the complex energy plane. The phase variation observed was always rather weak.

\begin{figure}[t]
  \includegraphics[height=.24\textheight]{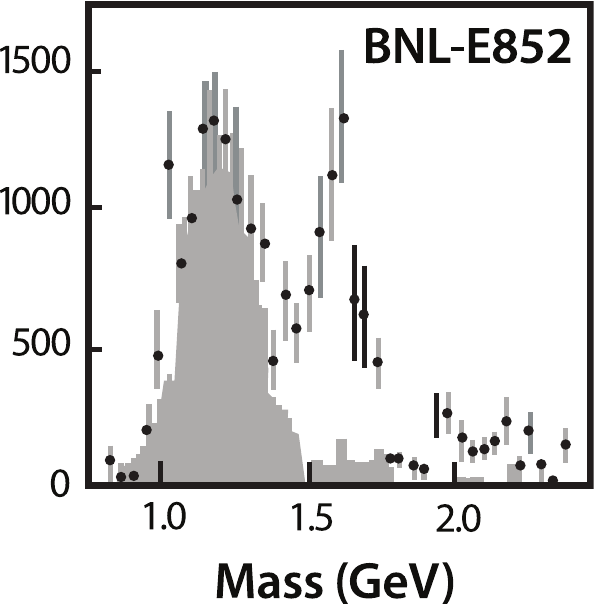}
 \hspace{5mm} \includegraphics[height=.245\textheight]{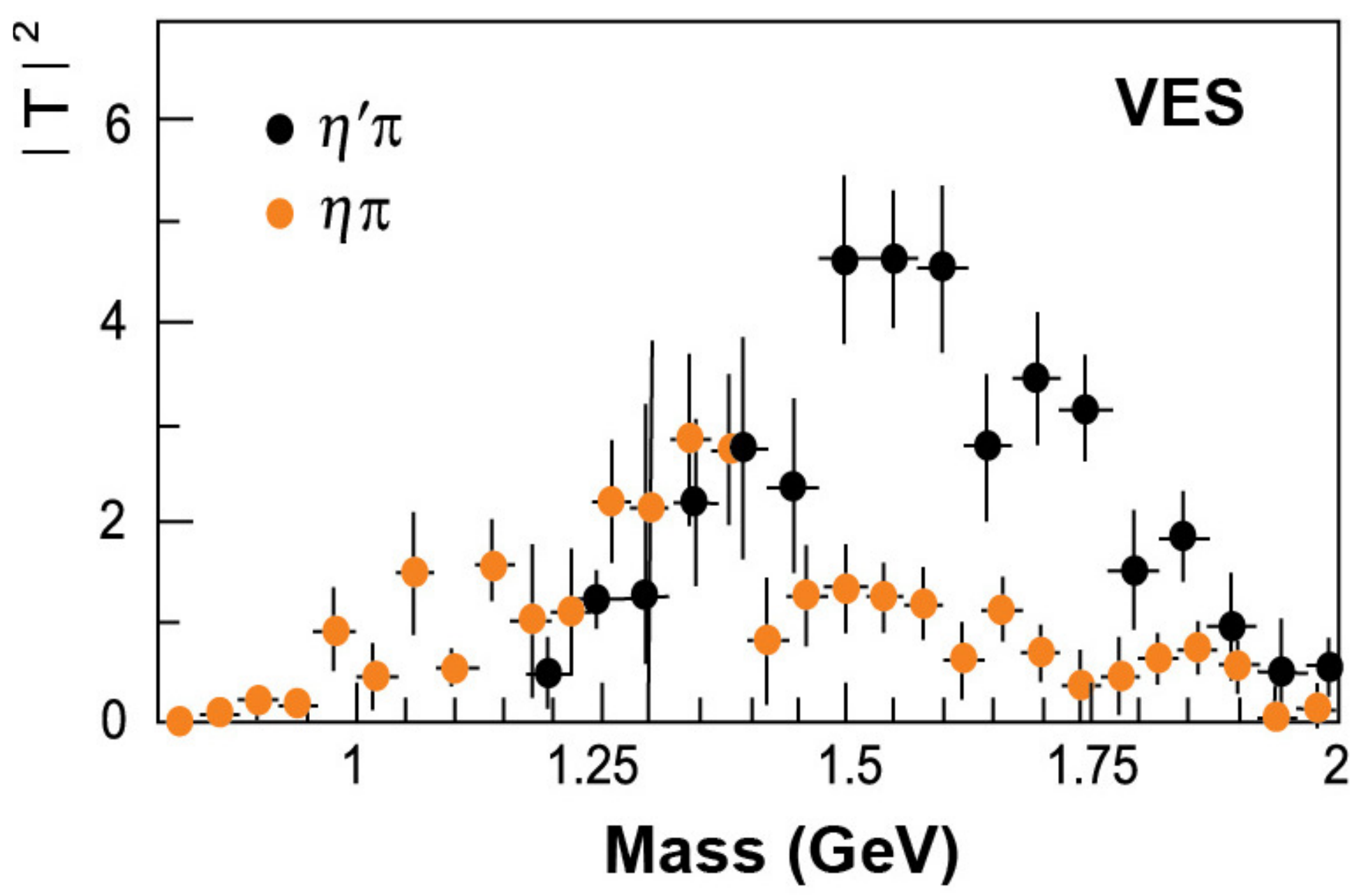}
  \caption{On the left is the $J^{PC}\,=\,1^{-+}$ signal from BNL-E852 data~\cite{chung} on $\pi N\to (3\pi)N$. The grey histogram is the calculated \lq\lq leakage'' into this channel from other partial waves. The enhancement at $\sim 1.4$ GeV is thereby explained~\cite{dzierba}, but leaves a clean $\sim 1.6$ GeV enhancement. The graph on the right displays the VES results~\cite{ves} on $\,\eta\pi\,$ and $\,\eta'\pi\,$ production as a function of the di-meson mass in $\,\pi^- Be\,$ collsions at 28 GeV$/c$, again with enhancements at 1.4 and 1.6 GeV, respectively. Whether any of these is resonant is unclear.}
\end{figure}

 A much more ambitious program is that of COMPASS@CERN. This studies multi-hadron production at small momentum transfers with a 192 GeV pion beam on nucleon and nuclear targets, in particular studying $\pi\eta'$ and $3\pi$ final states. The $\pi\eta'$ data show a significant broad enhancement in  $1^{-+}$ waves around 1600 MeV, but with little relative phase variation compared with the reference $2^{++}$ wave with its pronounced (conventional ${\overline q}q$) $\,a_2(1320)$ signal~\cite{compass1}. In the $3\pi$ channel, the first runs in 2004 showed a very crude enhancement in $1^{-+}$ waves, which was fitted to a Breit-Wigner form with doubtful significance~\cite{compass0}. However, now COMPASS are studying 96 million events in the $3\pi$ channel. With these statistics, one has to have a good understanding of the reaction mechanisms involved: simple Pomeron exchange with possibly important Deck effect backgrounds.
At last report the data require at least 52 partial waves to obtain a stable set. Only the dominant $2^{++}$ and $1^{+-}$ waves have been shown in talks. This meeting will elaborate more on this~\cite{compass2}. However, further work is needed to establish that there really is a $1^{-+}$ hybrid to add to the spectrum of physical hadrons.

A complementary effort is underway at Jefferson Lab with the instalation of magnets to increase the CEBAF energy to 12~GeV, a photon beam line and new detectors. A prime motivation for this upgrade is the search for hybrid mesons in all their quantum numbers, $J^{PC}$ and flavor: not just $1^{-+}$, but the $0^{+-}$, $2^{+-}$, {\it etc.}, expected at higher mass (Fig.~4).  GlueX is the new detector dedicated to studying multi-hadron final states created by an 11 GeV polarized photon beam on a proton target~\cite{gluex}. This is due to start taking data in 2016. Statistics comparable to COMPASS are expected, {\it i.e.} ~$10^8$ events. With wonderful angular coverage, this should allow small partial waves to be disentangled. Complementary (and occasionally competing) data on the low multiplicity final states will be taken by the CLAS12 detector at JLab too. 

The task of extracting small signals with certainty is a real challenge to experiment, phenomenology and theory. One most go beyond the simple isobar picture that was good enough, when one had even $10^4$ events. However, in the era of precision data one needs precision analyses too. This demands detailed knowledge of the reaction mechanisms involved, and importantly all the contributing final state interactions of $\pi$'s, $K$'s and $N$'s to be well-represented in terms of amplitudes that respect all the key properties of scattering theory. This requires a pooling of world expertise on partial wave analyses and $S$-matrix technology to ensure multichannel unitarity is fulfilled~\cite{ASI}. We have to learn from the experience of EBAC, Bonn-Gatchina, COMPASS and others, working with all the relevant analysis and experimental groups in the world. This will not just underpin the effort at JLab, but the same technology is required for comprehensive analyses of BESIII, LHCb and PANDA data. Steps are under way to bring this together. It is only by such collective efforts that we can be sure that signals of hybrids at the few percent level can be reliably extracted, and the poles of the $S$-matrix determined. It is not enough to confirm some putative $\pi_1(1600)$ signal (suggested by VES and BNL-E852), we must find the whole multiplet structure. It is only then that we can know that such \lq\lq exotic'' states are really hybrids of quarks and glue,
and not states with additional ${\overline q}q$ pairs, or hadronic molecules. The flavor multiplet structure is the guide~\cite{bali}. An understanding of the role of glue in QCD is the prize.    
   
Unless some real surprises happen, these experiments are likely to be the last in light hadron spectroscopy. If we are going to claim a real understanding of the detailed consequences of confinement, we had better get this right. That is the challenge for the next 10-15 years.
 
%%%%%%%%%%%%%%%%%%%%%%%%%%%%%%%%%%%%%%%%%%%%%%%%
%% BACKMATTER
%%%%%%%%%%%%%%%%%%%%%%%%%%%%%%%%%%%%%%%%%%%%%%%%

\begin{theacknowledgments}
  It is pleasure to thank the CIPANP organizers, particularly Wim van Oers and Martin Comyn,  for inviting me to give this talk.
The work was authored in part by Jefferson Science Associates, LLC under U.S. DOE Contract No. DE-AC05-06OR23177. 
\end{theacknowledgments}



\bibliographystyle{aipproc}   % if natbib is available

%\bibliography{sample}

\begin{thebibliography}{9}
\bibitem{bali} To answer a question from Gunnar Bali at this conference.
\IfFileExists{\jobname.bbl}{}
 {\typeout{}
  \typeout{******************************************}
  \typeout{** Please run "bibtex \jobname" to optain}
  \typeout{** the bibliography and then re-run LaTeX}
  \typeout{** twice to fix the references!}
  \typeout{******************************************}
  \typeout{}
 }
\end{thebibliography}

\end{document}